# Acoustic Localization Phenomena in Ferroelectric Nanophononic Devices

A. Bruchhausen[1,2,*], N. D. Lanzillotti-Kimura[1,3,*], B. Jusserand[4], A. Soukiassian[5],
D. G. Schlom[5,6], T. Dekorsy[2], A. Fainstein[1]

[1] Centro Atómico Bariloche and Instituto Balseiro (CNEA) and CONICET, 8400 S. C. de Bariloche, R. N., Argentina
[2] Center for Applied Photonics and Department of Physics, University of Konstanz, D-78457 Konstanz, Germany
[3] CNRS - Laboratoire de Photonique et de Nanostructures, Université Paris-Saclay, Route de Nozay, 91460 Marcoussis, France
[4] Institut des NanoSciences de Paris, UMR 7588 C.N.R.S.–Université Pierre et Marie Curie, 75015 Paris, France
[5] Department of Materials Science and Engineering, Cornell University, Ithaca, New York 14853, USA
[6] Kavli Institute at Cornell for Nanoscale Science, Ithaca, New York 14853, USA

[*] Equally contributors

*Abstract.*

*The engineering of phononic resonances in ferroelectric structures appears as a new knob in the design and realization of novel multifunctional devices. In this work we experimentally study phononic resonators based on insulating ($BaTiO_3$, $SrTiO_3$) and metallic ($SrRuO_3$) oxides. We experimentally demonstrate the confinement of acoustic waves in the 100 GHz frequency range in a phonon nanocavity, the time and spatial beatings resulting from the coupling of two different hybrid nanocavities forming an acoustic molecule, and the direct measurement of Bloch-like oscillations of acoustic phonons in a system formed by 10 coupled resonators. By means of coherent phonon generation techniques we study the phonon dynamics directly in the time-domain. The metallic $SrRuO_3$ introduces a local phonon generator and transducer that allows for the spatial, spectral and time-domain monitoring of the complex generated waves. Our results introduce ferroelectric cavity systems as a new tool for the study of complex wave localization phenomena at the nanoscale.*

The study of complex localization phenomena requires direct access to spatial, temporal and spectral information, which is not accessible in photonics and extremely challenging in electronics. In this work we unlock a new toolbox in the field of localization phenomena by using ferroelectric-based acoustic nanocavities as building blocks of complex structures.

Acoustic nanocavities[1–4] that confine and enhance phononic fields in the giga- and terahertz spectral range have been a subject of intense investigation for over a decade. The possibility of actually using acoustic phonons to control and modulate other excitations in solids, to process information, or to simulate other systems[5–9] has become the main motivation to investigate acoustic-phonon devices. An acoustic nanocavity is usually formed by two distributed Bragg reflectors (DBRs) enclosing an acoustic spacer[3], working in a similar way to how a classical Fabry Perot interferometer works for light. Layers with thickness of a few nanometers, and atomically flat interfaces are two stringent requirements to fabricate devices capable of manipulating acoustic phonons in the GHz-THz range. A plethora of acoustic phonon devices have been experimentally demonstrated, ranging from acoustic mirrors and optimized filters[10,11] to complex coupled-cavity structures[9]. Conventional semiconductors have usually been at the base of these devices mainly due to experience and maturity gained over the past few decades in the growth of such materials for electronics and optoelectronics applications. Nanocavities grown with other materials, such as ferroelectric and multifunctional oxides, have been less studied mainly due to the challenges of synthetizing heterostructures with the needed quality and precision. Despite this limitation, the superior acoustic performance and strong coupling between phonons, charge and light of these materials has been recently demonstrated in acoustic superlattices, nanocavities and other devices grown by reactive molecular beam epitaxy (MBE)[1,12–14]. The use of ferroelectric materials in nanophononic applications paves the way towards novel devices where new interactions and transduction mechanisms can be engineered.

Acoustic phonons in the GHz-THz range lack commercial sources like the ones used for audible sound. Coherent acoustic-phonon generation techniques usually require the use of pulsed lasers and a light/hypersound transducer.[15,16] The engineering and development of these transducers remains a grand challenge in nanophononics. Metals like gold, nickel and aluminum present excellent transduction capabilities in the near infrared (NIR) range[2,4,17,18] due to the photothermal coupling associated to large absorption coefficients. Conventional semiconductors and piezoelectric materials bring additional transduction mechanisms into the game,[4,19–23] and expand the application domains of nanophononics towards real electronic and optoelectronic devices. Interestingly, conventional semiconductor materials do not present the best possible phononic performance. $SrTiO_3$ (STO) has many properties that make it an attractive material for multifunctional-based material applications: it is a quantum paraelectric, it becomes superconducting at low temperatures (when highly doped), it can be used as a substrate for other multifunctional perovskites, it can be conductive or insulating depending on the dopants, etc. $BaTiO_3$ (BTO) is a standard ferroelectric with a Curie temperature that can be tuned up to 800 K by strain engineering.[24] BTO and STO are perovskite oxides that have optical gaps in the 350 nm wavelength range, and thus, they are transparent in the near-infrared spectral range.[13] As such, they have a limited potential to be used as transducers using standard coherent phonon generation techniques. $SrRuO_3$ (SRO), on the other hand, is a metallic perovskite with a strong absorption of light in the NIR spectral range, enabling a strong and localized generation and detection of longitudinal acoustic phonons, with the added value that can be epitaxially integrated with the aforementioned materials.

In this work we show the selective spatially localized generation and detection of acoustic phonons in a cavity based on a BTO/STO distributed Bragg reflector. We achieve spatially selective phonon generation and sensing by using SRO as the core layer of the cavity. Using high-speed asynchronous optical sampling (ASOPS) pump-probe spectroscopy,[25–27] cavity dynamics can be clearly analyzed directly within the time-domain. We also study the phonon dynamics of a coupled acoustic cavity system (the phononic equivalent of a hydrogen molecule),[5] where one of the cavity spacers was made of SRO. We report the direct observation of the time and spatial beatings in such an acoustic molecule. The generation of strain in one of the cavities, and the consequent exchange of energy with the second cavity can be clearly observed. With these two simple structures we set the building blocks for the study of complex localization phenomena, such as Anderson localization[28,29] and Bloch oscillations[30].

Bloch oscillations[30] are one of the most striking quantum localization effects in solid state physics: an electron in a crystalline potential subject to a constant force will oscillate instead of being uniformly accelerated. Besides electronics,[31,32] Bloch oscillations have been reported in a series of systems including molecular systems,[33] atoms in optical lattices,[34] photonic[35] and nanophononic[9] coupled cavity systems. By coupling a series of acoustic cavities, and introducing a gradient in the resonant frequencies, it is possible to mimic the effect in a nanophononic system.[5,36] We demonstrate the scalability of the structures studied by extending the use of SRO as a local transducer to a system composed of 10 coupled acoustic nanocavities[5] in order to excite and measure Bloch-like oscillations of acoustic phonons in the 100 GHz range.

The paper is organized as follows: Section II is devoted to the study of a single hybrid cavity; Section III introduces the concept of a hybrid phononic molecule and presents the experimental study of the phonon dynamics in these structures. Results on Bloch oscillations of acoustic phonons in coupled cavity systems are presented in Section IV. Finally, in Section V we conclude and present future perspectives.

## II. HYBRID ACOUSTIC NANOCAVITIES

In this section we introduce a device formed by a hybrid metal cavity (SRO) with a BTO/STO epitaxial oxide phonon mirror. A schematic of the structure is shown in Fig. 1 (panel a). The effective $\lambda/2$ cavity spacer is formed by a bilayer of SRO and STO. In this cavity structure, an acoustic mirror is formed by a phononic distributed Bragg reflector made of 6.5 bilayers of $BTO_{29}/STO_{49}$ where the subscripts indicate the number of unit cells of each material. The sample/air interface corresponds to a free surface boundary condition, resulting in total acoustic reflection, thus completing the cavity structure. Panel c of Fig. 1 contains the calculated surface displacement as a function of the acoustic phonon frequency for vibrations incident from the substrate side with unit amplitude. The acoustic bandgap is indicated with a double arrow. At the center of the stopband a confined mode can be distinguished, marked with a star. The atomic displacement profile inside the structure corresponding to the confined acoustic mode is shown in panel b of Fig. 1, where an exponential decay of the displacement amplitude is observed along the extension of the DBR.

The described hybrid nanocavity was grown on a $TiO_2$-terminated (001) STO substrate by reactive MBE. The sample was experimentally studied by reflection-type pump-probe experiments at room-temperature.[16] We used a high-speed asynchronous optical sampling (ASOPS) setup.[25–27] The ASOPS system is based on two femtosecond titanium:sapphire oscillators each providing 60 fs pulses at a f=0.8 GHz repetition rate. The repetition rates are stabilized at a fixed frequency difference $\Delta f$=5 kHz. For pump-probe experiments, one laser delivers the pump pulse–generating coherent longitudinal-acoustic phonons– while the second one acts as a probe–sensing the instantaneous optical reflectivity of the sample–, sampling the 1.25-ns-long time window. The temporal length of both pump and probe pulses at the sample position is about 200 fs due to uncompensated dispersion of the beam-shaping optics. The pump and probe wavelengths are 785 and 830 nm, respectively. It is important to highlight that at these wavelengths both pump and probe beams only interact with the SRO layer, implying a spatially selective transduction. In other words, the phonons are generated and detected exclusively in the SRO layer. A typical incident pump (probe) laser power of 100mW (10mW) was used.

The top panel of Fig. 2 shows the transient optical reflectivity measured (black curves) on the hybrid nanocavity. This curve was obtained by subtracting the slowly varying components induced by the excited carriers and the temperature evolution of the sample. A single-frequency oscillation decaying in time is observable. As we will discuss later, these oscillations correspond to the acoustic phonons confined in the SRO/STO cavity spacer. The red line corresponds to a simulation of the pump probe experiment based on a photoelastic interaction using an implementation of the transfer matrix method. For this simulation we assumed that: i) light absorption only occurs in the SRO and ii) the photoelastic constant is non-zero only in the SRO. The simulations reproduce the time decaying behavior of the monochromatic signal remarkably well. By performing a numerical Fourier transform of the extracted oscillations it is possible to recover the frequencies of the phonons modulating the optical properties of the SRO layer. The right panel of Fig. 2a shows the Fourier transform amplitude where one clear peak is present in the spectrum. The peak at ~96 GHz corresponds to the confined modes in the SRO cavity spacer. The red curve corresponds to the Fourier transform of the simulated signal in the left panel. The full width at half maximum of this cavity peak is ~2 GHz, corresponding to an acoustic quality factor of 50, determined mainly by the transmission to the substrate, as verified from comparison with the simulated spectra.

We have performed similar experiments on a nanocavity with the same sequence of layers, but without the SRO layer. In this case no coherent phonons at the infrared laser wavelengths used were observed due to the lack of an efficient local transducer. This confirms the important role of the SRO layer as a sub-THz phonon transducer compatible with high-quality epitaxial growth. It is also worth mentioning that an additional peak corresponding to the Brillouin mode exists at ~31 GHz (not shown), mainly originating in the massive substrate and aided by the very small residual optical transmission.

## III. DYNAMICS IN PHONONIC MOLECULES

By coupling two identical cavities it is possible to form the equivalent of an acoustic molecule. This structure has been investigated theoretically.[5] The experimental study, however, of the phonon dynamics in this structure has remained, to the best of our knowledge, elusive. In the phononic molecule the modes of the two individual cavities hybridize to form two eigenstates of the full structure: one symmetric and one antisymmetric with respect to the center of the molecule. Two aspects characterize the acoustic molecule: i) the presence of these two acoustic modes; and ii) an elastic energy distribution of these two modes that is shared by the two cavity spacers. In other words, when the system is excited at one of its resonant energies, there is atomic displacement in both cavity spacers. By the same token, when one individual cavity spacer is excited, and since both modes have projections onto this cavity spacer region, the two eigenmodes of the molecule will be excited with a characteristic beating that is a fingerprint of a coupled-mode. As we show below, the presence of the SRO layer allows for the localization of the phonon wave packet both in time and space.

In the phonon molecule structure studied one of the cavities is a hybrid acoustic cavity similar to the one studied in the precedent section, while the second degenerate cavity is an oxide BTO/STO distributed Bragg reflector with a $\lambda/2$ BTO spacer. A schematic of the sample is shown in Fig. 3a. The full structure consists on a SRO/STO cavity spacer, one DBR, a BTO cavity spacer and a second DBR, all epitaxially grown on a $DyScTiO_3$(110) substrate. Due to the high acoustic impedance mismatch between the BTO and the other two materials, only one period in the central DBR suffices to achieve a moderate coupling between the cavities. Again, the most important feature is that the generation and detection of the phononic effects is exclusively localized in the top cavity spacer, i.e., in the SRO cavity layer.

The calculated surface displacement for unit amplitude vibrations incident from the substrate as a function of energy is shown in Fig.3d. The double arrow indicates the extension of the acoustic minigap while the stars mark the spectral position of the resonant frequencies. Two clear modes at 85.9 and 101.5 GHz, symmetrically split from the uncoupled confined energy value 93.7 GHz, are observed inside the acoustic stopband. In Fig. 3b we show the simulated displacement profile corresponding to the two molecule eigenmodes. Note that the amplitude presents its maximum in the area of the acoustic cavity spacers and exponentially decays in the DBR towards the substrate. The different symmetry of the modes (symmetric or antisymmetric) is evident in the region of the second acoustic spacer (BTO). The simultaneous excitation of the two modes results in a displacement profile where the elastic energy is mainly concentrated in one of the cavity spacers. The blue (red) curve in Fig. 3c presents the elastic energy distribution resulting from the addition (subtraction) of the eigenmode displacement profiles where the energy is mainly localized in the hybrid SRO+STO (or conversely the BTO) layer.

The transient reflectivity curve measured on the hybrid molecule is shown in Fig. 2b. A clear beating signal is observable, with a period of 64.1 ps. This signal decays in time as in the case of the single cavity structure studied in the previous section. The periodic signal corresponds to a phonon package that spatially oscillates between the two cavity spacers, modulating the optical reflectivity of the SRO each time it reaches the surface cavity layer. It is the presence of this localized transducer that allows the launch, and the locally probing of the wave-packet dynamics. In red we present simulations of the pump-probe experiments that again reproduce remarkably well the phonon dynamics in the molecule. The Fourier transform of the measured (simulated) signal is shown in the right panel of Fig. 2b in black (red). Two clear peaks at 85.9 and 101.5 GHz are observable, which is the characteristic feature of the phononic molecule.

Figure 4 shows a simulation based on the transfer matrix method of the total acoustic displacement as a function of the spatial position (z) and time (t) corresponding to a localized generation in the SRO spacer at t=0. We first study the case of the simple hybrid single cavity (Fig. 4a). After the first few picoseconds, where the displacement is only localized in the SRO layer, it is possible to distinguish the propagating pulse-like component escaping towards the substrate. The phonons with the cavity mode frequency will remain trapped in the SRO/STO bilayer. A slow decay can be observed in the intensity of these phonons. Note that in the pump-probe experiments only phonons present in the SRO layer produce a modulation of the optical reflectivity at the cavity mode frequency. The vertical white line indicates the interface between the substrate and the structure. The small change in the slope of the escaping beam originates from the different effective speed of sound of the structure and the substrate. At this interface, due to that acoustic impedance mismatch, part of the acoustic beam is reflected towards the air/sample interface.

The phonon dynamics in the molecule are analyzed in Fig. 4b. As in the case of the single hybrid cavity, we can identify the propagating pulse-like component escaping towards the substrate. The remaining phonons show characteristic molecule beatings: the weight of the signal oscillates between the two spacers (located around 10 nm and 80 nm from the surface) with a period of 64.1 ps. The horizontal arrows match the time of those displayed in Fig. 3b and correspond to the situation where the "pure" symmetric (green arrow) or anti-symmetric (cyan arrow) mode combination is reached. By observing the acoustic displacement intensity at the top cavity layer it is possible to infer the periodic nature of the measured signal, demonstrating the importance of having a localized acoustic-phonon generator and detector in a complex structure. In this case, the SRO allowed us to excite and measure the beatings of the molecule using NIR pulsed lasers.

## IV. BLOCH-LIKE OSCILLATIONS OF ACOUSTIC PHONONS

Bloch oscillations of a particle take place under the action of a periodic potential when a constant force is acting on it. The application of a constant force results in the acceleration of the particle until it reaches the edge of the first Brillouin zone. At this point Bragg scattering of the particle by the lattice potential takes place, flipping the sign of the particle velocity. In the Brillouin zone, the particle reappears in the other edge of the first Brillouin zone with a negative velocity. In real space, under a constant force the resulting motion of the particle is oscillatory instead of uniformly accelerated. This phenomenon was originally proposed by Bloch and Zener for electrons,[30,37] and took some 60 years to be experimentally demonstrated.[31,32] In solid state physics the Wannier-Stark ladder is the spectral counterpart of Bloch oscillations, and is characterized by a series of equidistant eigenmodes.

The lack of charge of photons and phonons makes it difficult to define a photonic or phononic potential, respectively. A few years ago it was demonstrated that by concatenating a series of identical cavities it is possible to define a band.[5] The artificial band is centered on the energy of the individual cavities, the bandwidth is determined by the inter-cavity coupling, and two bandgaps are created (lower and higher energy bandgaps). It is also possible to mimic the effect of a potential by slightly detuning the energy of the individual cavities as a function of the position. The effect is equivalent to tilting the bands and bandgaps. This strategy led to the experimental demonstration of Bloch oscillations of photons, phonons and surface acoustic waves, among others.[9,35,36,38] This information is essentially concentrated in the observation and measurement of the Wannier-Stark ladder. The direct access to spatial localization information and to the oscillation dynamics of phononic Bloch oscillations is, however, still missing. By scaling-up the engineering of coupled acoustic cavities, and by using the unique characteristics of the SRO as a local generator and detector of coherent acoustic phonons, we now show that it is possible to spatially localize the measurement of Bloch-like oscillations.

We present both measurements and theoretical simulations of a system composed of 10 coupled acoustic cavity structures supporting Bloch-like spatial oscillations of acoustic phonons. Each cavity consists of a $\lambda/2$ spacer and a single $\lambda/4$ BTO layer acting as interference mirror. The first cavity spacer is made of SRO+STO, while the other nine spacers are made of STO. The structure was designed to present an energy gradient of ~10 GHz/cavity, starting with the SRO+STO cavity at 86 GHz. As in the cases of the single hybrid cavity and the phononic molecule, the sample was grown by reactive MBE on a $DyScO_3$ (110) substrate.

Figure 5b shows the surface displacement profile as a function of the phonon energy. The rich spectrum shows regular peaks between 60 GHz and 110 GHz, and a second series of peaks between 140 GHz and 220 GHz. Each peak corresponds to a phononic mode with the energy mainly localized between the air-sample interface, which acts as a perfect mirror and the region in space where coherent Bragg interference takes place. Note that this effect can be described in terms of tilted effective bands for the acoustic phonons as discussed in the supplementary information.

The measured optical transient reflectivity signal is shown in black in Fig. 5d. The slowly varying signal corresponding to the thermal and electronic evolution of the metal was removed for clarity. Well defined oscillations with decaying amplitude in time can be observed. The decay in amplitude is mainly attributed to transmission to the substrate as in the previous cases. The SRO layer provides a unique mean to measure time-resolved localized strains, allowing us, to measure the oscillations of complex acoustic phonon waves in the time domain. This same structure is also a local phonon generator, launching the acoustic phonons from a particular element in the array of phononic resonators.

Fourier transforms of both the measured (black) and simulated (red) signals are shown in Fig. 5c. A rich and complex Stark-ladder-like spectrum can be observed, and most of the measured peaks can be associated to the surface displacement peaks of Fig. 5b. The agreement between the simulated and the measured spectra is remarkable. These are Wannier-Stark ladder-like modes.

When the modes of the Wannier-Stark ladder-like are excited from the SRO cavity, the acoustic phonons will propagate until they reach the region on the sample matching a Bragg condition where they will be reflected, changing the sign of the phonon velocity. When the phonons reach the surface, they are reflected changing the sign of the oscillation. This perfect reflection can be seen as antisymmetric boundary condition: in this way, the full Bloch-like oscillation will be completed with a second round trip to the Bragg-condition zone of the sample. In Fig. 5d, the red and green lines are a guide to the eye to indicate the Bloch-like behavior of the detected signal, with a period of approximately ~65 ps. Each maximum corresponds to the arrival time of the phonon wavepacket to the SRO layer, the only place where they can be detected. Also noteworthy is the inversion of the signal that occurs in each detected period (indicated with green lines in Figure 5d), in full agreement with the total reflection suffered by the wavepacket at the surface of the sample. Once again, the role of the SRO layer is essential, acting as a spatially selective generator and detector of acoustic phonons, allowing for the spectral and time-domain monitoring of the complex wave phenomena.

## V. CONCLUSIONS

Using picosecond ultra-high frequency acoustics and local transducers we have shown that it is possible to access the full information characterizing complex wave localization phenomena, a unique feature as compared with photons and electrons. Additionally, GHz acoustic phonons, in addition, provide a unique way to couple to other relevant excitations in solids, for example to spins, two level systems, and polaritons, all relevant to novel quantum information technologies.
By employing a surface SRO cavity, we generated confined acoustic phonons. The metallic character of the SRO layer allowed us to selectively generate and monitor spatially localized acoustic phonons in the technologically relevant 100-GHz frequency range, paving the way to study novel materials in nanophononics. This structure was used as a building block to design a double cavity system, where a surface SRO cavity was coupled to a BTO cavity, forming a phononic molecule. The selective spatial excitation of confined acoustic phonons allowed us to experimentally localize the generated complex waves and thus study the coupling dynamics between the two acoustic-phonon resonators. Phonons were generated exclusively in the SRO cavity, and then the energy was transferred from the SRO and the BTO cavities. The beating dynamics of the two modes of the molecule has been clearly observed in the time domain. In the last section of this work we introduced a system formed by one hybrid acoustic nanocavity coupled to an array of nine cavities. Such a system is able to support Bloch-like acoustic-phonon oscillations. We experimentally showed that phononic modes trapped between the air-sample interface and the bandgaps resulting from the coupling of the nanocavity modes can generate spatial oscillations of the acoustic phonons when excited from the metallic surface. The direct access to spatial and time-domain dynamics unveils for the first time the full wave function behavior in these resonator-based structures. The potential of incorporating SRO as a light/hypersound transducer inside STO/BTO structures thus constitutes one of the first steps towards the engineering of new nano-optophononic applications involving complex structures and the study of localized light-charge-phonon interactions.

**ACKNOWLEDGMENTS**

A.E.B. acknowledges the Fellowship of the Alexander von Humboldt Foundation (Germany). N.D.L.-K. acknowledges support from the ERC through the MC-IIF OMSiQuD. A.S. and D.G.S. gratefully acknowledge the financial support from the National Science Foundation through the MRSEC program (DMR-1420620). This work was performed in part at the Cornell NanoScale Facility, a member of the National Nanotechnology Coordinated Infrastructure (NNCI), which is supported by the National Science Foundation (Grant ECCS-1542081). T.D. acknowledges support through the DFG (SFB 767).

**AUTHOR CONTRIBUTIONS**

A.F. and N.D.L.-K. proposed the concept. A.S. and D.S. fabricated the samples. A.B. and T.D. conducted the experiments. A.B., N.D.L.-K. and A.F. performed the theoretical simulations, data analysis and wrote the manuscript. All the authors discussed the results. AF guided the research.

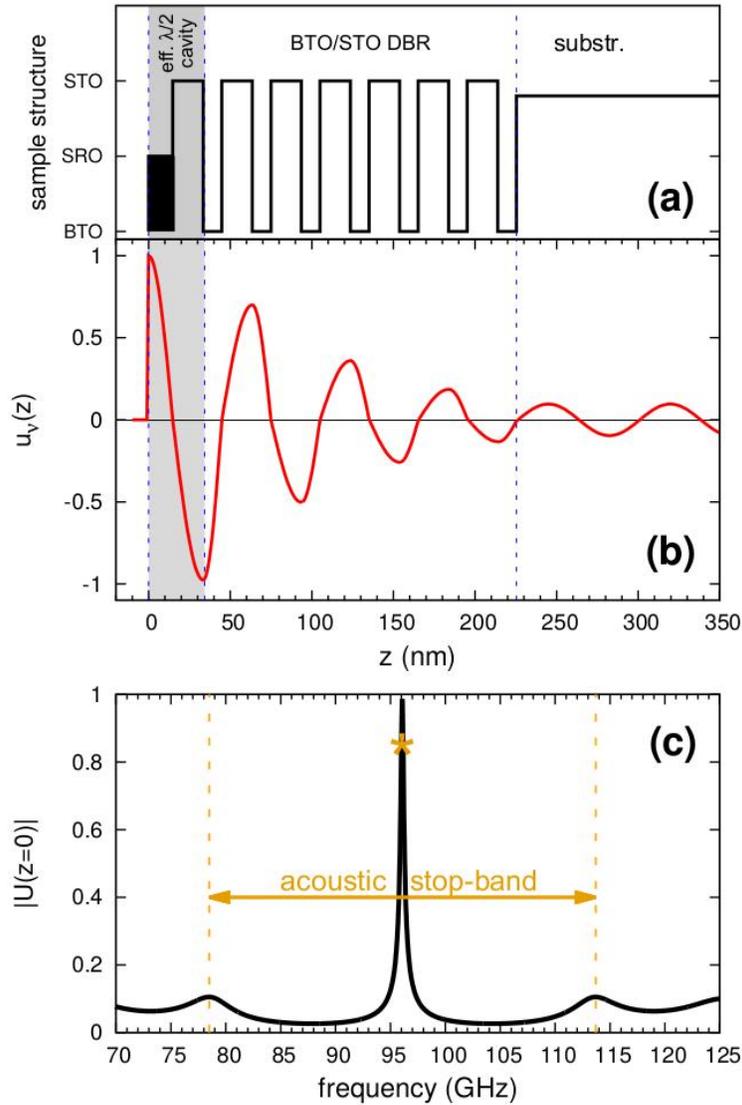

**Figure 1.** Hybrid acoustic nanocavity structure. (a) Schematics of the structure formed by a SRO acoustic spacer on top of a STO/BTO based DBR. (b) Calculated displacement profile corresponding to the resonant mode. The maximum of the field is located at the top bilayer (SRO+ STO). (c) Calculated surface displacement as a function of frequency; the double arrow indicates the extension of the acoustic minigap while the star marks the position of the resonant frequency.

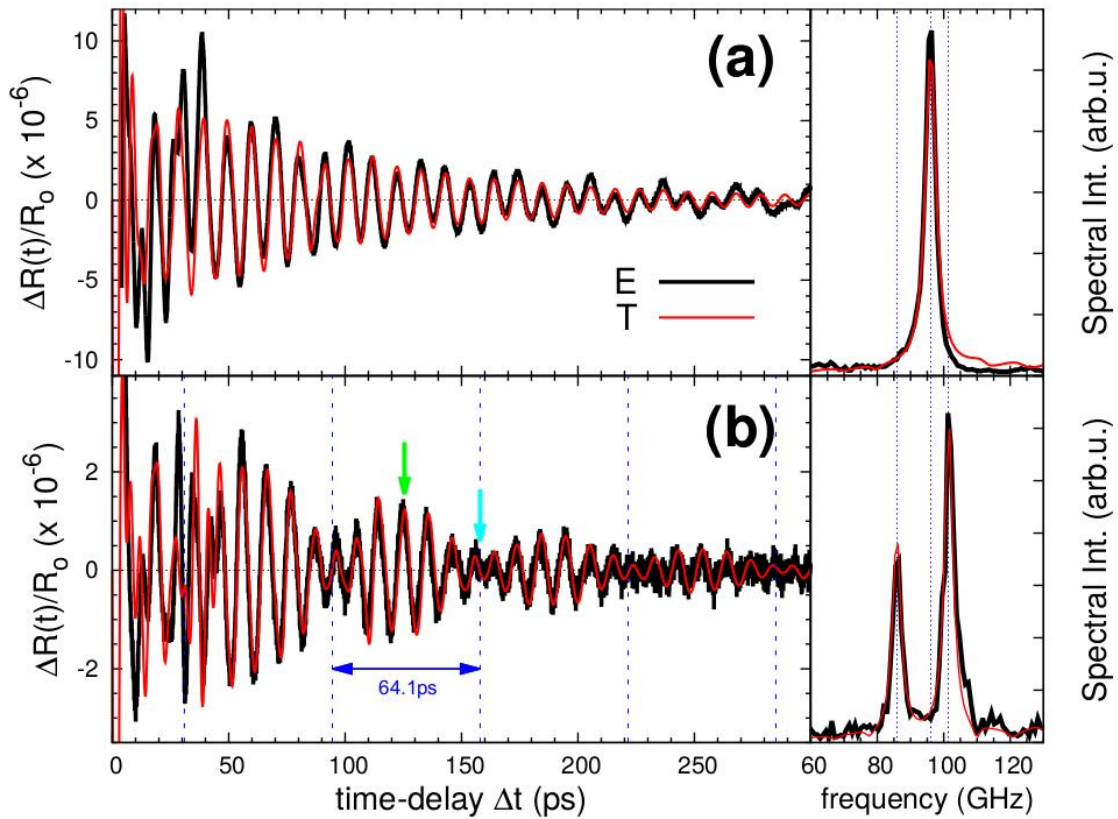

**Figure 2.** Optical transient reflectivity of the hybrid phononic cavity (a) and molecule (b). Left panels present the measured (black) and simulated (red) signals using the ASOPS system. Slowly varying components of the signals related to the electronic and thermal evolution of the sample were subtracted for clarity. Right panels show the Fourier transform of the measured and simulated time signals.

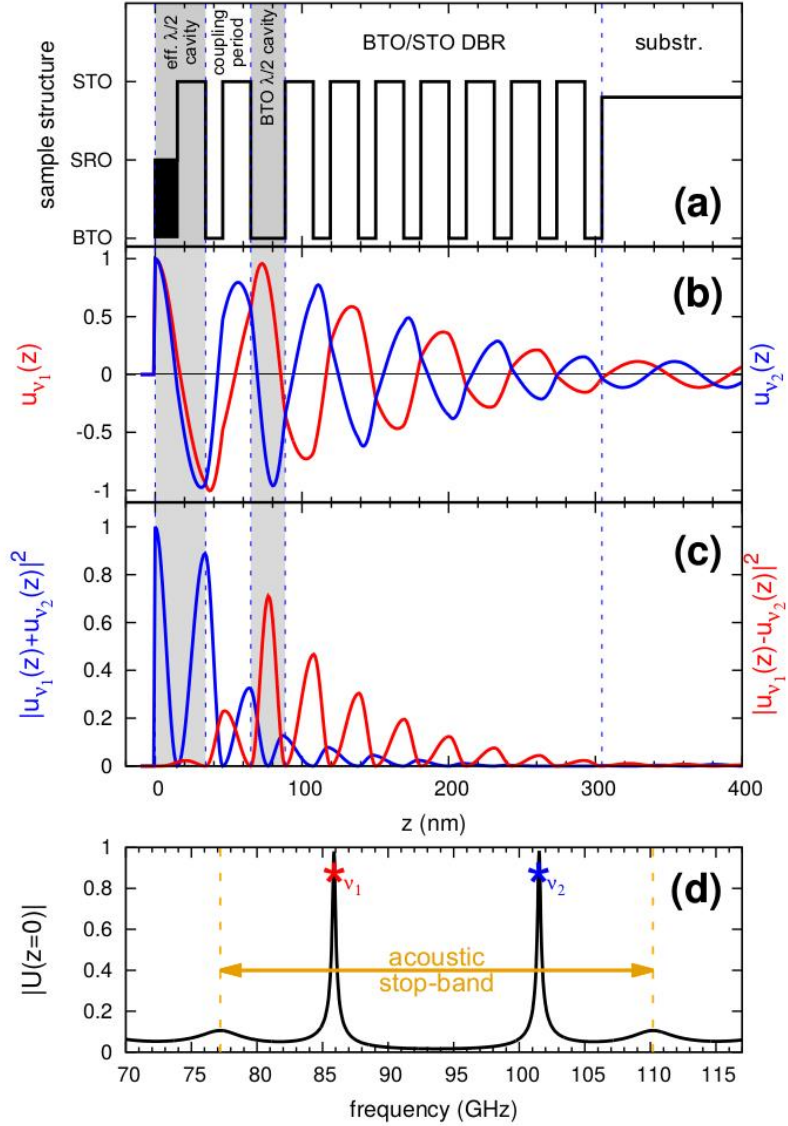

**Figure 3.** Hybrid acoustic molecule. (a) Schematics of the phononic molecule formed by an air/spacer/DBR cavity and a DBR/spacer/DBR cavity. (b) Calculated displacement profile of the two resonant modes. Observe that the weight of the displacement is mainly located around the two molecule spacers. (c) Elastic energy distribution corresponding to the addition and subtraction of the two eigenmodes of the molecule. The blue (red) curve denotes the energy localized in the hybrid SRO+STO (BTO) layer. (d) Calculated surface displacement as a function of energy; the double arrow indicates the extension of the acoustic minigap, while the stars mark the spectral position of the resonant frequencies.

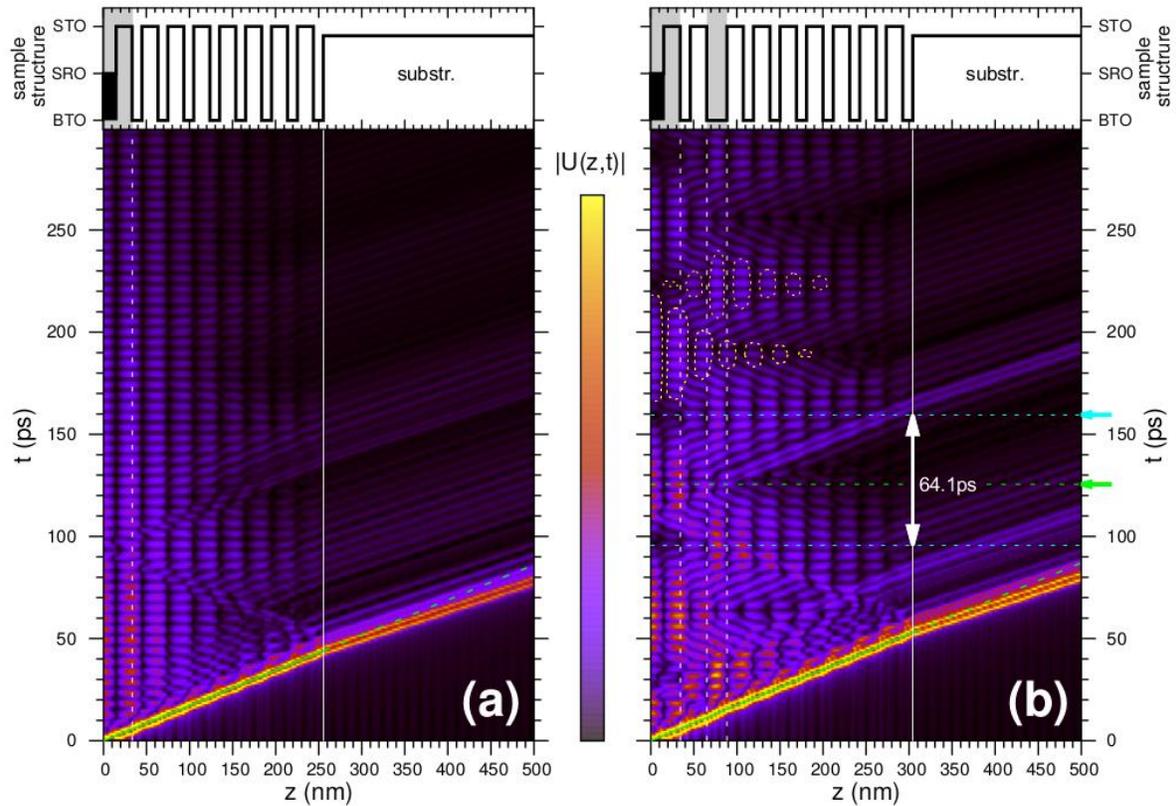

**Figure 4.** Intensity map of the acoustic displacement in the hybrid devices as a function of the spatial position (z) and time (t) corresponding to a localized generation in the SRO spacer at t=0. Brighter regions correspond to larger displacements. (a) Hybrid cavity structure case. (b) Phononic molecule case. The elastic energy oscillates between the two cavity spacers. The vertical double arrow indicates the period of the oscillation, while the dotted areas are guides to the eyes of the spatial distribution of the two localized modes.

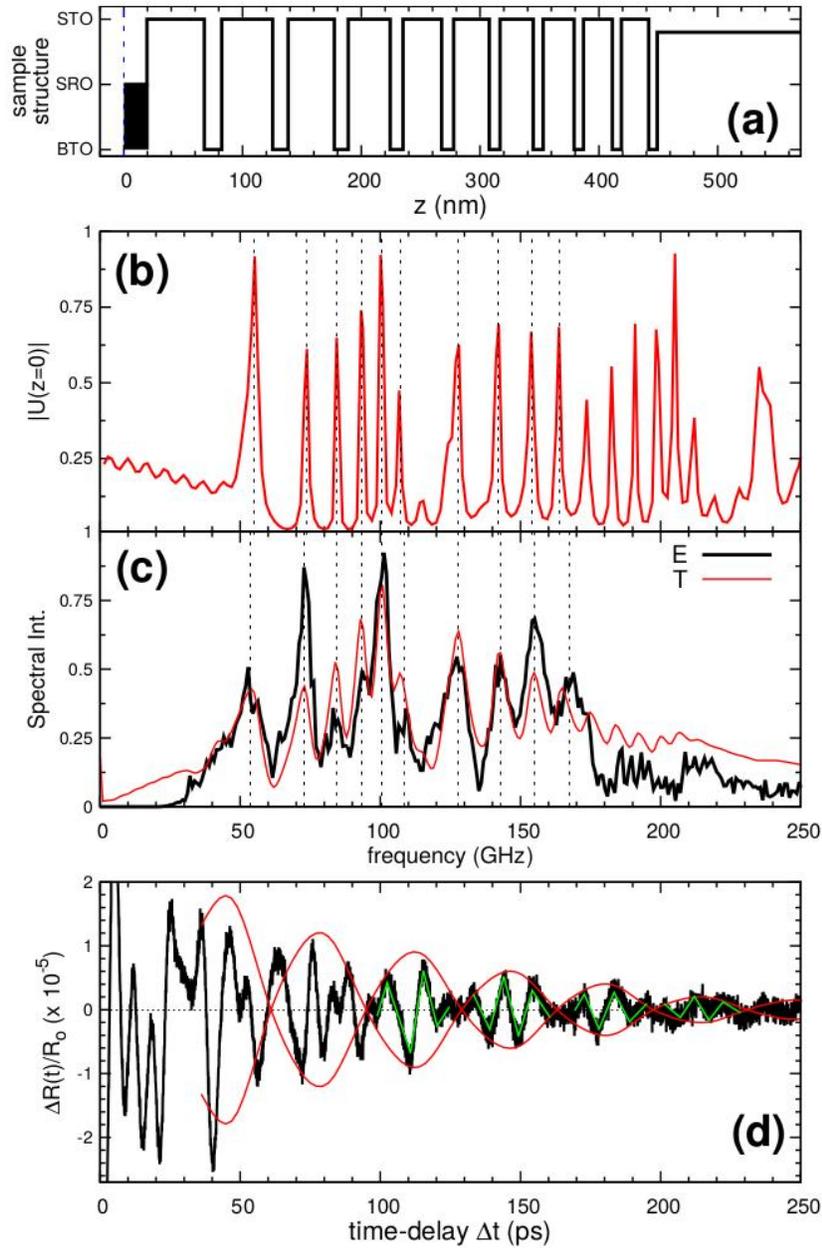

**Figure 5.** Bloch oscillations of acoustic phonons in hybrid complex structures. (a) Schematics of the phononic multiple-cavity resonator formed by a series of concatenated cavities with a gradient in the individual resonant energy. The first cavity is an air/spacer/DBR cavity. (b) Surface displacement as a function of the frequency. Note the series of peaks corresponding to the acoustic Wannier-Stark ladder. (c) Fourier transforms of the measured (black) and simulated (red) optical transient reflectivity signals. (d) Measured (black) optical transient reflectivity signal. The red lines are a guide to the eye to indicate the oscillations corresponding to the acoustic wavepacket reaching the SRO layers.